\documentstyle[emulateapj]{article}

\slugcomment{To appear in ApJ Letters}

\lefthead{\"Ostlin}
\righthead{HST/Nicmos imaging of IZw18 }

\begin{document}

\title{HST/NICMOS Observations of IZw18 -- revealing a population of
old AGB stars\footnote{Based on observations
with the NASA/ESA {\it Hubble Space Telescope}, 
obtained at the Space Telescope Science Institute, which is operated by the 
association of Universities for Research in Astronomy, Inc., under NASA 
contract NAS5-26555.}}

\author{G. \"Ostlin\altaffilmark{2}}
\affil{Institute d'Astrophysique de Paris, 98bis Boulevard Arago, 
75014 Paris, France}


\altaffiltext{2}{Present address: Stockholm Observatory, 133~36 Saltsj\"obaden, 
Sweden  (email: ostlin@astro.su.se)}


\begin{abstract}
I present the first results from a HST/NICMOS imaging study of the 
most metal-poor blue compact dwarf  galaxy IZw18.
The near infrared color--magnitude diagram (CMD) is dominated by two populations, 
one 10-20 Myr population of red super giants and one 0.5-5 Gyr population
of asymptotic giant branch  stars.  
Stars older than 1 Gyr are required to explain the observed
CMD at the adopted distance of 12.6 Mpc, showing that IZw18 
is {\emph {not}} a young galaxy. The results hold also if the distance to IZw18 is significantly 
larger. This rules out the possibility that IZw18 is a truly young galaxy
formed recently in the local universe.

\end{abstract}




%

\section{Introduction}

The blue compact galaxy (BCG) IZw18 (also known as Markarian 116) was first described 
by Zwicky in 1966. In a seminal paper Searle and Sargent (\cite{ss}) showed that IZw18 had a 
very low oxygen abundance, and they concluded that the galaxy either must be young
or form stars in short bursts intervened by long quiescent epochs (cf. Searle et al.
\cite{ssb}). This has been followed by many studies of the chemical properties of
IZw18 and BCGs in general. Alloin et al. (\cite{alloin}) derived an oxygen abundance 
of 12+log(O/H)=7.2 (2\% solar) for IZw18 in agreement with more recent values, e.g. 
Izotov  and Thuan (\cite{it}). In spite of many searches for more metal-poor
galaxies, IZw18 remains the most metal-poor one (when nebular oxygen abundances are 
considered) and  is often regarded as the prototypical BCG (see Kunth and 
\"Ostlin \cite{ko} for a review).

If IZw18 was a genuinely young galaxy, that would imply that galaxies can form still
at the present cosmic epoch and considerable effort has therefore been devoted to 
examine the 
ages of BCGs.
Although it is now clear that most BCGs are old (see Kunth and \"Ostlin \cite{ko}), 
IZw18 has remained among a small number of galaxies which show no clear signs
for the presence of an  old stellar population. 
 Kunth and Sargent
(\cite{ks}) argued that IZw18 could be young and that the low, but non-zero, 
metallicity could be due to self-pollution by massive stars. 
Recently an extensive study of chemical abundances of BCGs was published by 
Izotov and Thuan (\cite{it}). They found that for very low oxygen abundances 
(less than 5\% solar) the N/O and C/O ratios stay constant and have very little scatter. 
Their interpretation is that N and C are produced entirely by massive ($M \ge 10 M_{\odot}$)
stars and that this implies that these galaxies must be younger than 40 Myr. While other 
interpretations are possible (Kunth and \"Ostlin \cite{ko}), this an interesting idea 
that deserves to be tested. With its record low oxygen abundance, IZw18 provides the
critical test case.

The chemical abundances of IZw18 have been used to infer both a high and a low 
age (Kunth and \"Ostlin \cite{ko}), and it is clear that chemistry alone 
has limited value in
constraining the age and star formation history  given the many uncertain 
physical parameters involved (winds, yields, infall, initial mass function, etc.). 
Stronger constraints would be obtained by direct observations of the resolved 
stellar population.
Hunter and Thronson (\cite{hunter:thronson}) used deep HST/WFPC2 imaging to 
investigate the stellar population of IZw18, and found a population of young 
massive stars.
No evidence for old stars was found, but neither any against. Similar results 
were obtained  by Dufour et al. (\cite{dufour}) from an independent WFPC2 data set.
Lately, these two  WFPC2  data sets have been carefully reanalyzed by Aloisi, Tosi and 
Greggio (\cite{aloisi}), who discovered a population of red stars. Their results indicate that 
IZw18 is at least 0.5 Gyr old. This was however disputed by Izotov et al. (\cite{izotov}) 
who claim that the distance to
IZw18 has been underestimated, and that hence the age has been overestimated.

In this paper I report on the first results from a near infrared imaging
study of IZw18 using NICMOS  on board the 
HST. The goal was to search for an old stellar population, by identifying 
luminous red giants and asymptotic giant branch (AGB) stars. The data is however also 
valuable for studying the 
young massive stars, e.g. red super giants (RSGs) in the near infrared (NIR). A follow 
up program designed at investigating the presence of carbon stars by targeting molecular
features were 
performed during {\sl  Cycle 7-NICMOS}. The results of this investigation
will be presented in a forthcoming paper.

\section{Observations and Reductions}

I Zw 18 was observed for 3 orbits during {\sl Cycle\,7} with NICMOS in the F110W 
and F160W passbands (roughly corresponding to J and H) under GO program 7461. The 
instrument and its performance are described  in Thompson et al. (\cite{thompson}) 
and the NICMOS instrument handbook (MacKenty et al.  \cite{mackenty}).
The NIC2 camera was chosen to give good pixel sampling (0.075 $\arcsec$/pixel) 
and an acceptable
field of view ($19\arcsec \times 19\arcsec$).
All science 
exposures were obtained with the MULTIACCUM mode which employs consecutive 
non-destructive readouts of the detector. This enables e.g. an efficient 
cosmic-ray removal. 
For the F110W filter, two MIF1024 sequences were acquired, separated by an 
offset of 10 integer pixels in the Y direction. In addition, one MIF512 sequence was 
obtained with a 0.5 pixel offset in both the X and Y directions with respect 
to the first MIF1024 sequence. The use of a non-integer pixel offset 
was adopted  since the NIC2 pixel scale slightly undersamples
the point spread function (PSF) in F110W. 
For the F160W filter, two sets of images, consisting of one MIF512 and 
two MIF1024 sequences, were acquired at the same positions as the two MIF1024 
sequences in F110W. The total exposure times are thus 2560 and 5120 seconds 
in the F110W and F160W filters respectively.

\begin{figure}
\resizebox{\hsize}{!}{\includegraphics{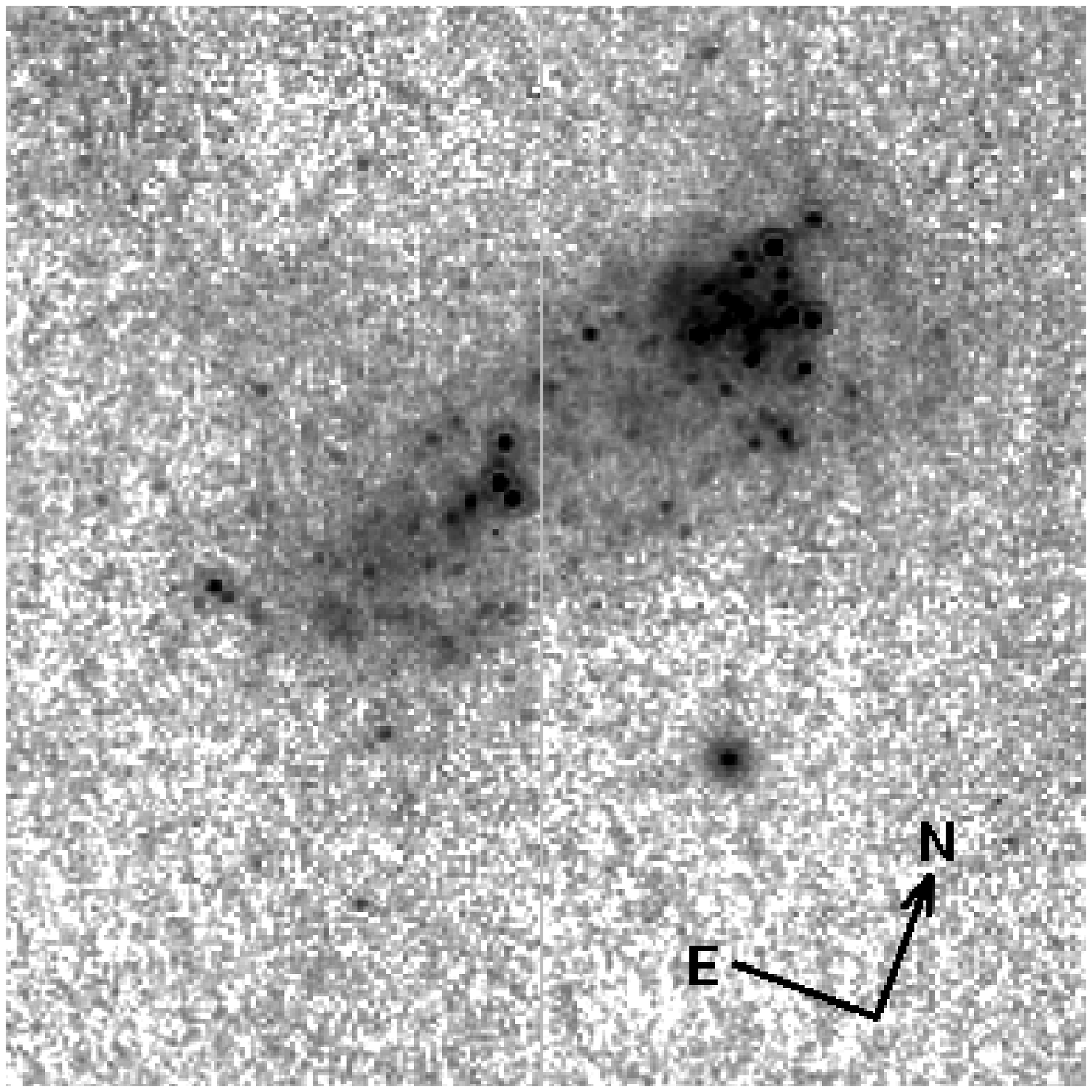}}
\caption{The central $14.5\arcsec \times 14.5\arcsec$ of the drizzled NICMOS F160W image 
of IZw18. The north and east directions are indicated 
at the lower right. The length of the arrow is $2\arcsec$.}
\label{fig1}
\end{figure}

Inspection of the pipeline processed images revealed that these suffered
from the so called ``pedestal'' effect, a residual bias not removed by the pipeline.
To subtract the pedestal, the  Pedestal Estimation and Quadrant Equalization 
Software developed by R.P. van der Marel was successfully utilized.
After the pedestal had been removed, the images in each pass band were combined using
the Drizzle package (Fruchter and Hook \cite{fruchter}). The resulting F160W image is shown 
in Fig. \ref{fig1}.

\placefigure{fig1}

Photometry of point sources was performed with the {\tt daophot} package
(Stetson et al. \cite{stetson}). Unfortunately, the images contain
few suitable stars for modeling the point spread function (PSF), since  the  
bright stars are found in crowded regions.  One bright source 4\arcsec \ south-west 
(SW) of the main body turned out to be spatially extended and is probably a background 
galaxy.
To assure that the wings of the PSF were not contaminated by neighboring stars,
photometry was also performed using a synthetic PSF, which  was constructed using the 
TinyTim  software (version 4.4, Krist and Hook \cite{krist}). The resulting photometry  
using real and synthetic PSF stars were nearly identical, except for a systematic 
difference of the order of 0.05 magnitudes in the F160W filter;  the use of synthetic 
PSF stars yielded redder stars.  Comparison with a control sample consisting of the eight 
best isolated stars showed that the synthetic PSF photometry did not suffer from any 
systematic offset compared to the aperture photometry, whereas the use of a natural PSF 
showed a small systematic offset but smaller scatter than the synthetic PSF photometry. 
Thus, the natural PSF photometry is probably more accurate but subject to a small 
systematic offset.  No systematic 
difference was seen 
in the F110W data. The natural PSF photometry was finally adopted after correcting for 
the small offset in F160W.
 After the photometric reduction, stars that resulted too sharp or too extended 
in at least one filter   were rejected. The photometries in the two 
filters were then combined by considering a matching radius of 0.5 pixels in X and Y for 
the stellar centers in the F110W and F160W images.
All objects were carefully analyzed by eye and dubious sources rejected. 

\begin{figure}
\resizebox{\hsize}{!}{\includegraphics{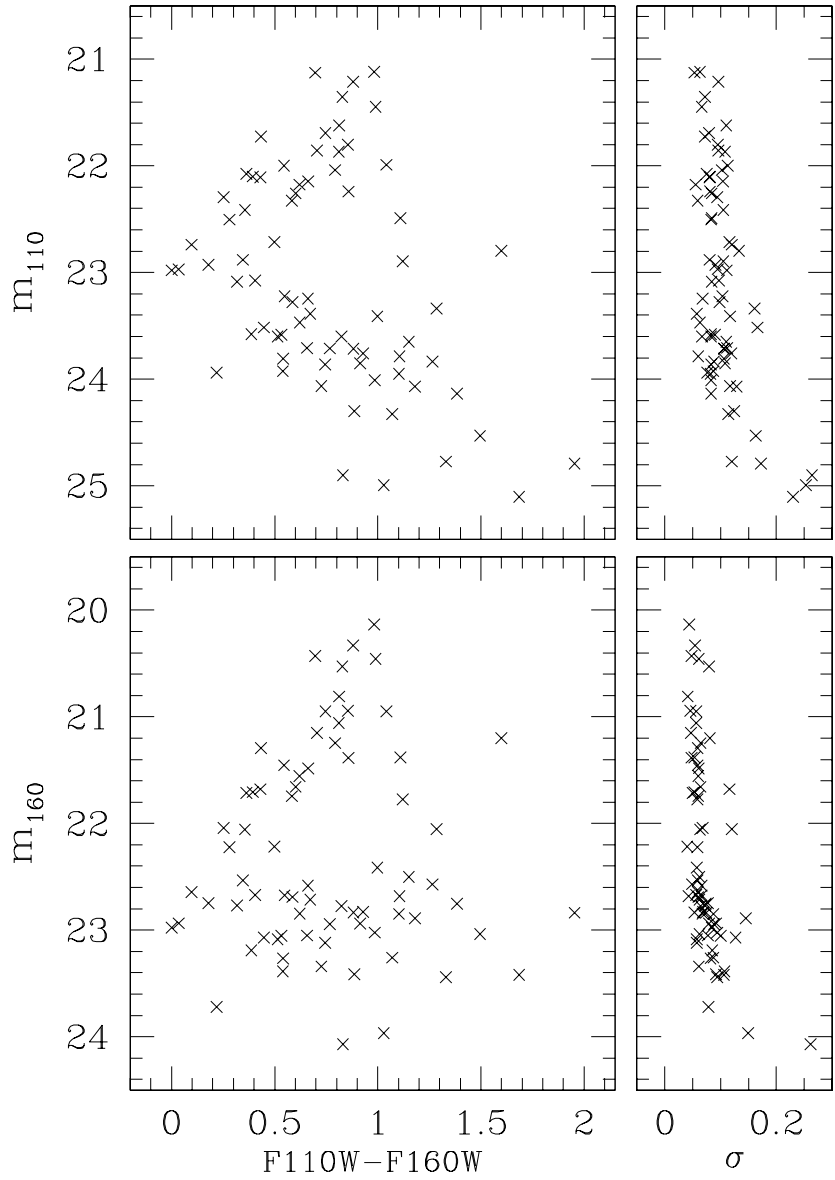}}
\caption{The left panels show the observed color--magnitude diagrams, and
the right panels show the observed magnitudes versus photometric uncertainties
for F110W (top) and F160W (bottom). }
\label{fig2}
\end{figure}

The DAOPHOT magnitudes refer to an aperture with radius 0.1875\arcsec . 
The aperture corrections in going from this to an  0.5\arcsec \ 
 aperture were determined using synthetic PSF and amount to 0.24 and 0.48
 magnitudes for F110W and F160W respectively.
To obtain total magnitudes an aperture correction of $-0.156$ magnitudes for 
both filters were added to the the 0.5\arcsec \ aperture magnitudes, 
and finally, the zeropoints of $ZP_{F110W}= 22.5436$ and $ZP_{F160W}= 
21.826$ were added in order to have the final calibrated magnitudes in the HST 
Vegamag system  (see NICMOS WWW pages).
I will use the designations $m_{110}$ and $m_{160}$ to express the magnitudes in 
 F110W and F160W respectively, and ${\rm F110W - F160W}$ for the color.
The resulting color--magnitude diagrams (CMDs) and the magnitude--error relations 
are shown in Fig. \ref{fig2}. The detected stars span a similar
range in color and brightness as those detected with HST/NICMOS in 
VIIZw403 (Schulte-Ladbeck et al. \cite{schulte}). 

\placefigure{fig2} 

The photometric completeness was investigated by using
the {\tt addstar} task. Artificial stars were added 
in half magnitude bins at random
locations in two separate regions: ``SE'' covering  the  mildly crowded 
south-east region; and ``NW'' which is the severely crowded  north-west  region. Then, 
repeated tests were made of how many test stars within each half magnitude bin that were 
successfully recovered by daophot. More than 100 stars were added for each bin.
The 75 and 50 \% completeness levels are 
summarized in Table 1.
In the NW region there are still bright unresolved
sources left after the repeated daophot runs. 

\placetable{table1}

\section{The distance of IZw18}

Usually a distance of 10 Mpc is assumed for IZw18, based on a radial velocity 
of $\sim$750 km/s and a Hubble constant of $H_0 =75$km/s/Mpc. Recently Izotov et al. 
(\cite{izotov}) have claimed that the distance to IZw18 has been severely
underestimated, and if Virgo-centric flow is taken into account a distance 
of 15-20 Mpc is obtained. They present also other circumstantial evidence
(the presence of Wolf-Rayet stars and the ionization state of IZw18) for a 
distance $D \sim 20$ Mpc. It is true that Virgo-centric motion will
slightly increase the distance, but the distance proposed by  Izotov et al. 
(\cite{izotov}) refer to old values of the distance  of the Virgo cluster (21.7 Mpc).
A lot of recent progress has been made on the distance to Virgo (e.g. Macri
et al. \cite{macri}) showing that its distance is $D_{Virgo} \approx 16$ Mpc.
The Virgo-centric flow models express the distance of a galaxy as $x$, relative to 
the Virgo cluster, i.e. $D = x \times D_{Virgo}$ . Applying the linear model 
by Schechter (\cite{schechter}) gives $x=0.7$ while the non-linear models by Kraan-Korteweg 
(\cite{kraan}) give $x=0.8$, corresponding to 11 and 13 Mpc respectively.
Hereafter I will adopt $D = 12.6$ Mpc for IZw18, which corresponds to a distance
modulus of $(m-M)=30.5$.  The distance should not be much larger (maximum 14.5 Mpc
given the Virgocentric flow models and the uncertainty in $D_{Virgo}$)
than this, but may be as low as $\sim 10$ Mpc. The distances 
suggested by  Izotov et al. (\cite{izotov}) would require IZw18 to be more 
distant than the Virgo cluster, in spite of a much lower radial velocity.

\section{Analysis and Discussion}

\begin{figure}
\resizebox{\hsize}{!}{\includegraphics{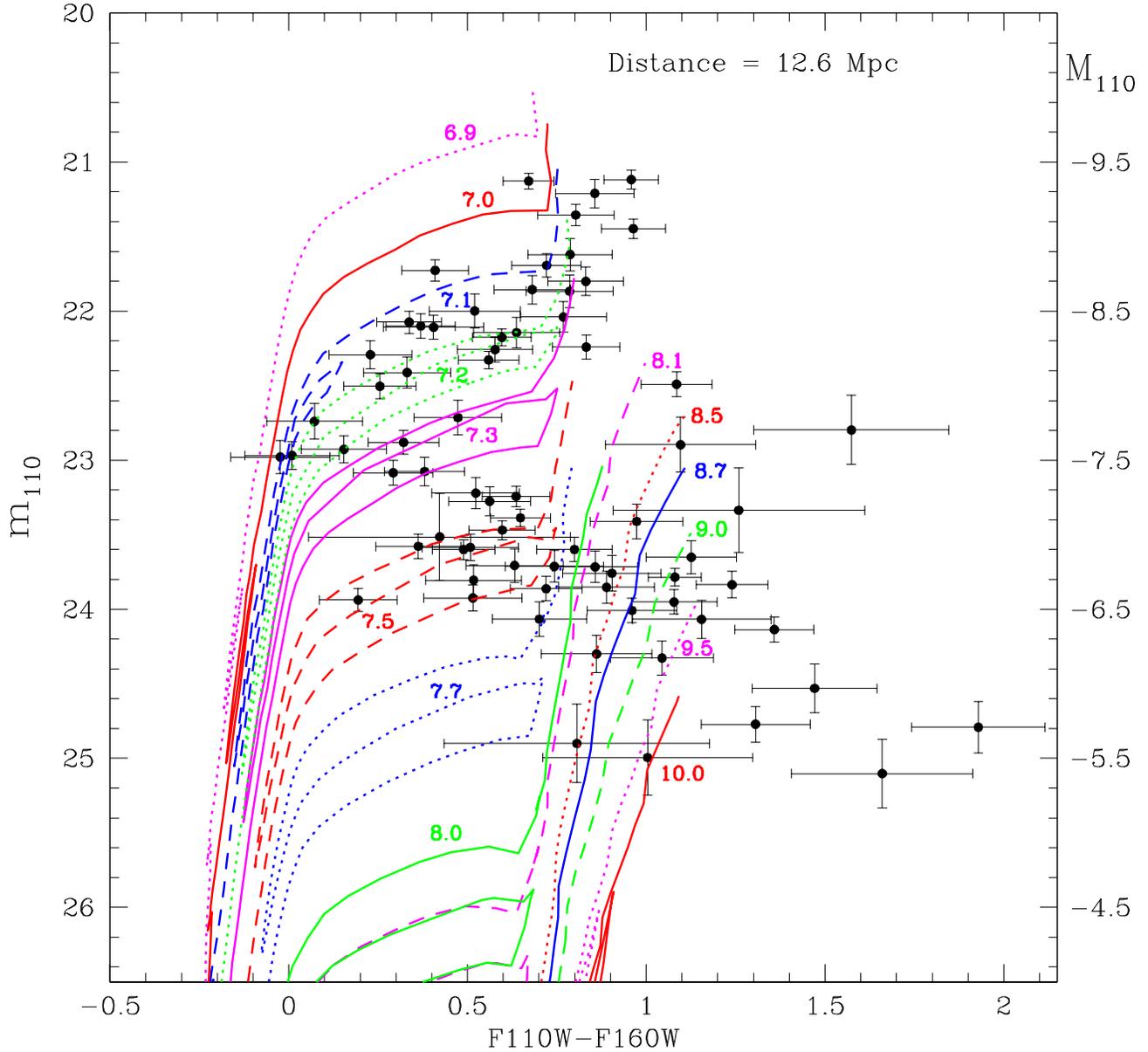}}
\caption{Observed color--magnitude diagram (corrected for extinction) with 
isochrones from Bertelli et al. (1994) over-plotted. The numbers correspond 
to the logarithm of the 
age for each isochrone.
The right hand
axis gives the absolute F110W magnitude for a distance of 12.6 Mpc.}
\label{fig3}
\end{figure}

Internal extinction has  only a moderate influence on the F110W-F160W colors
and in general $E({\rm F110W-F160W}) = 0.5 \times E(B-V)$, which has been estimated
with the Synphot software. All optical investigations have found a low internal 
extinction in IZw18. I have adopted $E(B-V)=0.05$ 
throughout the galaxy, and this correction has been applied to the data-points in 
Fig. \ref{fig3}.  

The obtained color magnitude diagram (CMD) clearly presents two distinct
features. One population of red luminous stars  and one with fainter 
($m_{110} > 23$) red stars. This suggests that stars with at least two 
different ages are present. To interpret the observed CMD I have mainly used the 
Padua stellar models for a metallicity $Z=0.02 Z_{\odot}$ (Fagotto et al. \cite{fagotto}, 
Bertelli et al. \cite{bertelli}), i.e. identical to the gas metallicity of IZw18.
In general, the Bertelli et al. (\cite{bertelli}) isochrones are more useful than
the Fagotto et al. (\cite{fagotto}) tracks since the former contains later evolutionary
stages, as the thermally pulsating AGB phase, which is very important for the present 
study.

Before interpreting the CMD, the effective temperature 
($T_{\rm eff}$) and luminosity ($L$) of the stellar models have to be 
related to colors and absolute magnitudes in the observed
system. To translate $T_{\rm eff}$ into  F110W-F160W colors, and $L$ 
into $M_{110}$ for a given surface gravity, the BCP models by Origlia 
and Leitherer (\cite{ol}) were used for a metallicity of $[\rm{Fe}/\rm{H}]=-2$. 

In figure \ref{fig3}
the Bertelli et al. (\cite{bertelli}) isochrones are over-plotted on the CMD. For
a distance modulus of $(m-M)=30.5$ it is obvious that stars of ages 11-20
Myr are required to reproduce the luminous upper branch in the CMD ($m_{110} < 23.2, ~
F110W-F160W < 1$). These ages equals the lifetime of stars with initial masses of
20 to 13 $M_\odot$. The upper branch is dominated with stars $\approx 14$ Myr old
when the current star formation event appears to 
have peaked. For comparison, distances of 10 and 15 Mpc would give ages of 18 and 12 
Myr respectively.   
 Comparison with the Fagotto et al. (\cite{fagotto}) stellar evolutionary tracks yield
the same conclusion that the most luminous red stars are the evolved descendants of 
stars with initial masses 15-20 $M_\odot$, and the same results are 
obtained when using ``Geneva'' tracks for the same metallicity (Schaerer 1999, private 
communication; cf. de Mello et al. \cite{demello}).

\placefigure{fig3}

For low mass stars (initial mass $M \la 5 M_\odot$, age $> 100$ Myr) only the isochrones 
(Bertelli  et al. \cite{bertelli}) have been considered, since the tracks do not include 
the thermally pulsating AGB phase. The most interesting result is that to reproduce 
the faint red part of the CMD ($m_{110} > 23.5, F110W-F160W > 1 $), an age of 1 Gyr is 
required at the distance of 12.6 Mpc.
Even if the distance would be as high as 15 Mpc, stars older than 0.5 Gyr would be 
required, and $D=10$ Mpc would lead to a much higher age than 1 Gyr. Not even at 20 Mpc 
can the faint red stars be made younger than 300 Myr. 
For $D=12.6$ Mpc the red faint population can be nicely explained
as the upper part of AGB stars with a range of ages from 0.3 to 5 Gyr.
This corresponds to the lifetime of stars with initial masses 2.5 to 1.1 $M_\odot$. 
A single 0.5 Gyr old AGB population does not suffice since  that  would not produce 
enough faint ($m_{110} \approx 24$) red stars, only when extending to ages in
excess of 1 Gyr can the CMD of faint stars be satisfactory explained.  This indicates that
IZw18 has experienced more or less continuous star formation for at least 1 Gyr prior to
the present burst.

The distribution of stars in the CMD is broadened by the observational
error function, and it has to be assured that it cannot be reproduced 
by young stars plus photometric errors. 
If one assumes that no star is older than 100 Myr, what distribution would one expect?
An isochrone of that age  has a color $F110W-F160W=0.79$ at $m_{110}=24$.
The one sigma photometric errors at $m_{110}=24, F110W-F160W\approx 1$ are 
$\sigma_{F110W-F160W} = 0.125$, but let's be very conservative 
and adopt $\sigma_{F110W-F160W} = 0.2$. 
The one sigma range around $F110W-F160W=0.79$, $m_{110} \sim 24$ then contain 14 stars.
For a normal distribution, 68\% of the stars should be contained within this range, 
and one would thus expect to find only $\sim 3$ stars (i.e. 16\%) with $F110W-F160W \ge 1.0$, 
wheras  one finds more than a dozen. Thus the observed distribution of red stars 
cannot be reproduced by 0.1 Gyr old stars plus photometric errors. Even if rejecting 
the reddest stars ($ > 1.5$),  an age of 1 Gyr is required to be consistent with the 
error distribution. Since this was a conservative estimate, we can conclude that IZw18 
has an age of at least 1 Gyr.
This result on the near-infrared CMD analysis is confirmed when using the synthetic
CMD technique (cf. Tosi et al. (\cite{tosi}); Aloisi et al. \cite{aloisi}): stars 
older than 1 Gyr are required in order 
to reproduce the observed CMD, the best fit  being obtained by including stars as old 
as 5 Gyr (\"Ostlin 2000, in  preparation).

An analysis of the spatial distribution of stars of different age show that young 
red super giants occupy preferentially the NW region, while the old AGB stars are
more evenly distributed.  
Further constraints on the age of the stellar population may be obtained from 
the color of the underlying galaxy after the found stars have been subtracted.
For a single stellar population, the integrated F110W-F160W
color increases with age, from $F110W-F160W=-0.1$  at 10 Myr up to  $F110W-F160W=0.6$
at 125 Myr when the AGB stars start to dominate, and stays roughly constant thereafter.
This has been calculated  using the PEGASE spectral synthesis code (Fioc \& 
Rocca-Volmerange 1999) and the NICMOS F110W and F160W response curves.
The integrated  colors in $15 \times 15$ pixels boxes
are typically $F110W-F160W = 0.3$ to $0.6$ in the NW and $F110W-F160W > 0.6$ 
in the SE region. 
This  indicates that in the NW region, the
light is dominated by an unresolved population of stars with ages 20 to 100 
Myr, while in the SW region, the underlying population is dominated by stars
older than 125 Myr.

The result of an age of 1 Gyr or larger
for the underlying population in IZw18 is confirmed also by  deep optical/
near-IR surface photometry (\"Ostlin 2000, in preparation; Kunth \& \"Ostlin 
\cite{ko}) and is in good agreement with the results of Aloisi et al. (\cite{aloisi})
for the optical CMDs. We can thus conclude that all these different studies 
on independent sets of data in various spectral regions give the same answer: 
that IZw18 is an old galaxy.

\section{Summary and Conclusions}

I have analyzed the near infrared color-magnitude diagram from HST/NICMOS
images of IZw18, the most metal-poor galaxy known. Assuming a  distance of 
12.6 Mpc, I find support for the presence of AGB stars 
 at least 1 Gyr old.  The results presented in this 
letter hold qualitatively even if the distance to IZw18 is significantly different. 
The latest star formation event 
appears to have peaked 14 Myr ago. My results are in good  agreement
with Aloisi et al. (\cite{aloisi}) who from a careful reanalysis of 
HST/WFPC2 archival data  concluded that IZw18 must be at least 0.5  Gyr old
at their adopted distance of 10 Mpc. 
Thus, there is no need for galaxies with oxygen abundance less
than 5 \% of the solar value to be young as was claimed by Izotov and Thuan 
(\cite{it}).
This closes the question whether IZw18 is a genuinely young galaxy, still in
the process of formation: the answer is -- No!

\acknowledgments

Thanks to D. Calzetti and W. Freudling for practical help and suggestions, and to D. Schaerer 
for making revised evolutionary tracks available ahead of 
publication. N. Bergvall, D. Kunth, A. Lan\c{c}on, M. Longhetti, M. Mouhcine 
and R. Schulte-Ladbeck are thanked for useful discussions.
This work made use of the Pedestal Estimation and 
Quadrant Equalization Software developed by Roeland  P. van der Marel.
The referee A. Aloisi is thanked for many useful suggestions.
This work was financially supported by the Swedish foundation for international
cooperation in research and higher education (STINT), and the Swedish National 
Space-board. Many thanks to Anna and Snigel.

\clearpage

\clearpage

\begin{table}
\begin{center}
\begin{tabular}{crrrrr}
\tableline
completeness & \multicolumn{2}{c}{F110W} & & \multicolumn{2}{c}{F160W} \\
level    &  SE & NW & & SE & NW \\
\tableline
75\%    &   23.6    &   22.8   &     & 22.6  &  22.0 \\
50\%    &   24.0    &   23.5   &     & 23.4  &  22.7 \\
\tableline
\end{tabular}
\end{center}
\tablenum{1}
\caption{Completeness limits in magnitudes, for the two regions discussed in the text.
\label{table1}}
\end{table}


\end{document}